\documentclass[preprint,12pt]{elsarticle}

\usepackage{amssymb}

\usepackage{lmodern}
\usepackage[T1]{fontenc}
\usepackage{xcolor}
\usepackage{placeins}
\usepackage{graphicx}
\usepackage{dcolumn}
\usepackage{mathptmx}
\usepackage{etoolbox}
\usepackage{multirow}
\usepackage{amsmath}
\newcolumntype{M}[1]{>{\centering\arraybackslash}m{#1}}
\definecolor{mycolor}{rgb}{1,0,0}
\definecolor{dgreen}{rgb}{0,0.75,0}

\journal{Chaos, Solitons \& Fractals}

\begin{document}

\begin{frontmatter}



\title{Polarization in the three-state $q$-voter model with anticonformity and bounded confidence.}


\author[inst1]{Arkadiusz Lipiecki}
\ead{lipiecki.arkadiusz@gmail.com}

\affiliation[inst1]{organization={Department of Theoretical Physics, Wrocław University of Science and Technology},
            addressline={Wybrzeże Wyspiańskiego 27}, 
            city={Wrocław},
            postcode={50-350},
            country={Poland}}
            
\author[inst1]{Katarzyna Sznajd-Weron\corref{cor}}
\ead{katarzyna.weron@pwr.edu.pl}

\cortext[cor]{Corresponding author}

\begin{abstract}
Engaging with dissenting views, fostering productive disagreements or strategic anticonformity can benefit organizations as it challenges the status quo. The question arises, however, whether such strategic anticonformity ultimately leads to social polarization, which is not a desirable phenomenon. We address this question within an agent-based model of discrete choices. Using the way of modeling social responses in continuous opinion models, we propose a three-state $q$-voter model with anticonformity and bounded confidence. We analyze the model on a complete graph using the mean-field approach and Monte Carlo simulations.  We show that strong polarization appears only for a small probability of anticonformity, which means that conformity combined with homophily enhances polarization. Our findings agree with results obtained previously in the group discussion experiment and within various continuous opinion models.
\end{abstract}

\begin{keyword}
agent-based model \sep discrete choice \sep polarization \sep opposing viewpoints \sep bounded confidence \sep voter model \sep Human Resource Management

\MSC 37Hxx \sep 37Mxx \sep 37Nxx
\end{keyword}

\end{frontmatter}


\section{Introduction}
Imagine that as a member of a certain organization you have to make a decision on the wildlife control method and you have to choose between three alternatives: (1) doing nothing, (2) the non-lethal control, and (3) the lethal control method. All other members of your organization have to make such a choice, and on this basis, you will give the recommendation to the government. Now imagine two scenarios for making this decision: in the first scenario you have to give an answer immediately, and in the second one the moment for making a decision is postponed for some longer time, and in the meantime you can repeatedly discuss with others. In this paper we focus on the second scenario, which means that we take into account social influence -- both positive (conformity) and negative (anticonformity) \cite{Nai:Dom:Mac:13}. 

If the discussion is to ultimately lead to a certain recommendation, polarization is definitely undesirable, especially since it may be an irreversible phenomenon \cite{Mac:etal:21}. As noted by M\"as and Flache, explanation of polarization often \textit{hinge on the assumption of negative influence} \cite{Mas:Fla:13}. Therefore it seems that anticonformity is an unfavorable social response during  determination of a common view. On the other hand, opposing viewpoints maybe valuable for the team performance \cite{Lan:Fou:Hol:18}. As noted recently by Minson and Gino \cite{Min:Gin:22}: \textit{Much has been written on the benefits for teams and organizations of engaging with opposing views (...) Yet anyone who has been involved in such work knows that disagreements on strongly held opinions, (...), are always tough and frequently destructive.}

The above observation is the direct motivation for this work. In this article, we ask whether anticonformity actually reinforces polarization. It was shown that polarization can be explained without negative influence, and may be induced by the positive influence combined with homophily \cite{Mas:Fla:13,Def:etal:02,Heg:Kra:02, Par:And:Mel:16, Maa:Dal:Wal:20}. Both conformity and 
homophily, which is perfectly described by the quote \textit{similarity breeds connection} \cite{McP:Smi:Coo:01}, are undeniably strong social forces. The question remains, if additionally also anticonformity appears in the system, will it strengthen or rather weaken polarization? It is this specific question that we will be answering in this work.

The three alternatives on the wildlife control method, mentioned at the beginning of this article, were actually used in the empirical studies \cite{Lio:etal:17}, but here are given just as an example of a choice between three ranked items - two extremes (1,3) and one central (2). It means that in our model agents will be described by the three-state variable, often called opinion. To model interactions between employees, we take into account two basic types of social response: conformity (positive influence) and anticonformity (negative influence) \cite{Nai:Dom:Mac:13}. Moreover, we use the idea of bounded confidence, which was originally introduced into models with continuous opinions \cite{Def:etal:02,Heg:Kra:02}. Such models are usually based on the assumption of positive social influence and bounded confidence can be treated as a special case of homophily \cite{Mas:Fla:13}: agents who have opinions that do not differ by more than a certain threshold interact with each other and then their opinions become even more similar. However, in several continuous opinion models also negative interactions were considered \cite{Sta:Tes:Sch:08, Bis:Cha:Sen:12, Juu:Por:19,Gra:Li:20}.

The idea that agents who are similar to each other are likely to interact and in result they become even more similar has been introduced already in 1997 by Axelrod \cite{Axe:97}. However, Axelrod did not introduce any threshold beyond which interactions are strictly forbidden. Moreover, state of an agent was not described by a single continuous opinion but by the vector of $F$ discrete variables, which can take one of $s$ integer values. In our model agents are described by only a single variable, which corresponds to $F=1$, that can take one of $s=3$ values. However, we use the concept of bounded confidence, as in continuous models, and additionally consider negative social influence, which is not present in the Axelrod model.  

We assume that the range of bounded confidence related to conformity may be different from that related to anticonformity. This assumption is based on the knowledge from social science, within which many factors reinforcing conformity has been identified \cite{Nai:Dom:Mac:13, Mar:Men:14,Bon:05,Nai:Mac:Lev:00}. For example, it is known that conformity is more likely under the influence of unanimous group. Therefore our model is based on the $q$-voter model (qVM), in which an agent (target) changes opinion under the influence of a unanimous group of $q$ agents (source) chosen at random from all of the target's neighbors \cite{Cas:Mun:Pas:09}. Moreover, it is well known that people conform much easier to others that are similar to them, which corresponds to the principle of homophily. It means that bounded confidence related to conformity should be rather small. Much less is known about anticonformity and hence our assumption of independence of bounded confidence ranges for conformity and anticonformity \cite{Mar:Men:14}. 

\section{\label{Model} Model description}
We consider the system of $N$ agents, placed in the nodes of a given graph, who we also refer to as voters, since our model is based on the $q$-voter model \cite{Cas:Mun:Pas:09}.
Each voter $x$ at time $t$ is assigned a dynamic variable $s_x(t)$:
\begin{equation}
s_x(t) = i, \quad x\in \{1,2,\ldots, N\}, \quad i\in \{1,2,3\},    
\end{equation}
representing the opinion of the agent $x$. We say that agents are neighbors if the vertices they occupy are adjacent, that is, if there is an edge connecting them. The opinion of a voter (target) can change as a result of the influence of a unanimous group of $q$ neighbors. If within such a group an opinion of at least one agent differs from others, then the target's opinion does not change. As in many other versions of the $q$-voter model, the influence group, also called the $q$-panel, the source of influence, or simply a source, is formed by drawing without repetition a group of $q$ neighbors from all neighbors of a target voter \cite{Jed:Szn:19}. It means that $q$ is the parameter of the model and does not define the structure of the graph. The only constraint on the value of $q$ is that $q$ cannot be greater than the minimum degree of a vertex in the graph. In this paper, we focus on the infinite complete graph (CG), which means that all agents are mutually connected, for two reasons: (1) it allows for an analytical treatment, since it is equivalent to the mean-field approach (MFA) and (2) it allows for the comparison with other versions of the multistate $q$-voter model, which so far have been considered only on CG \cite{Now:Sto:Szn:21,Now:Szn:22,Don:Lip:Szn:22}.

In the case of binary-state models, the implementation of conformity and anticonformity is relatively straightforward \cite{Jed:Szn:19}. If an agent acts as a conformist, they adopt the opinion of the influence group and, in case of anticonformity, takes the opposite one, rebelling against the source. However, for the multi-state opinion there are many other possibilities. For example, Nowak et al. proposed that in a case of conformity a target voter $x$ takes the opinion $s_q(t)$ of the $q$ panel, while in the case of anticonformity they take equally likely any opinion other than $s_q(t)$ \cite{Now:Szn:22}. However, according to many psychological models of social influence, conformity involves moving the opinion of the target toward that of the influence group, while 
anticonformity is a change that moves it away from the source of influence \cite{Nai:Dom:Mac:13}. In this paper, we decided to employ the latter approach, as it is more psychologically justified. Such a way of modeling social responses is also often used in continuous opinion models \cite{Def:etal:02,Gra:Li:20}.

We assume that within a single update of duration $\Delta t$, the agent can change their opinion by exactly one: $s_x(t+\Delta t) = s_x(t) \pm 1$. Furthermore, we implement the bounded confidence (BC) rule, which limits interactions to opinions that satisfy $|s_q(t) - s_x(t)| \leq R$ , where $R$ is the range of bounded confidence \cite{Def:etal:02,Heg:Kra:02}. We allow $R$ to be different for conformity, $R=R_c$, and anticonformity, $R=R_a$. According to the above assumptions, the elementary update is as follows:
\begin{enumerate}
\item randomly draw a target from a uniform distribution $x \sim U \{1,N\}$,
\item randomly draw $q$ neighbors of $x$ without repetition to form the $q$-panel,
\item if the $q$-panel is unanimous, denote its opinion by $s_q(t)$ and randomly draw the number from a uniform distribution $r\sim U(0,1)$
\begin{enumerate}
\item if $r < p$ then the target anticonforms to the $q$-panel if $|s_{x}-s_{q}| \le R_a$, which means that their opinion $s_x(t)$ is repelled from $s_q(t)$ by one unit; see Tab. \ref{tab:tran_anti}.
\item 
otherwise, the target conforms to the $q$-panel if $|s_{x}-s_{q}| \le R_c$, which means that their opinion $s_x(t)$ moves towards  $s_q(t)$ by one unit, see Tab \ref{tab:tran_con}.
\end{enumerate}
\end{enumerate}
A time unit, which corresponds to one Monte Carlo step (MCS), consists of $N$  elementary updates.

\begin{table*}[t]
    \centering
    \begin{tabular}{| M{2.5cm} | M{1.5cm} | M{1.5cm} | M{2.5cm} | M{2cm} |}
    \hline
        $|s_x(t)-s_q(t)|$ & $s_x(t)$ & $s_q(t)$ &  $s_x(t+\Delta t)$ & probability \\
         \hline
         \multirow{4}{*}{0} & 1 & 1 & 2 & 1\\
           & 3 & 3 & 2 & 1\\
           & 2 & 2 & 1 & 1/2\\
           & 2 & 2 & 3 & 1/2\\
         \hline
        \multirow{2}{*}{1} & 2 & 1 & 3 & 1\\
          & 2 & 3 & 1 & 1\\
         \hline
    \end{tabular}
    \caption{Possible transitions $s_x(t) \rightarrow s_x(t + \Delta t)$ in the case of anticonformity. The last column represents the conditional probabilities that a corresponding opinion change occurs, given that the target with the opinion $s_{x}(t)$ anticonforms to the source with the opinion $s_q(t)$. For $R_a=0$ only transitions for $|s_x(t)-s_q(t)|= 0$ are possible, for $R_a=1$ all transitions for $|s_x(t)-s_q(t)|= 0$ and $=1$ are possible. The case of $R_a=2$ is equivalent to $R_a=1$ because no other transitions are possible in the 3-state model.}
    \label{tab:tran_anti}
\end{table*}

\begin{table*}[t]
    \centering
    \begin{tabular}{| M{2.5cm} | M{1.5cm} | M{1.5cm} | M{2.5cm} | M{2cm} |}
    \hline
        $|s_x(t)-s_q(t)|$ & $s_x(t)$ & $s_q(t)$ &  $s_x(t+\Delta t)$ & probability \\
         \hline
         \multirow{4}{*}{1} & 1 & 2 & 2 & 1\\
           & 3 & 2 & 2 & 1\\
           & 2 & 1 & 1 & 1\\
           & 2 & 3 & 3 & 1\\
         \hline
        \multirow{2}{*}{2} & 1 & 3 & 2 & 1\\
          & 3 & 1 & 2 & 1\\
         \hline
    \end{tabular}
    \caption{Possible transitions $s_x(t) \rightarrow s_x(t + \Delta t)$ in the case of conformity. The last column represents the conditional probabilities that a corresponding opinion change occurs, given that the target with opinion $s_{x}(t)$ conforms to the source with opinion $s_q(t)$. For $R_c=0$ no transitions are possible. For $R_c=1$  transitions for $|s_x(t)-s_q(t)|= 1$ are possible, while for $R_c=2$ all transitions for $|s_x(t)-s_q(t)|= 1$ and $=2$ are possible.}
    \label{tab:tran_con}
\end{table*}

\section{Mean-field approximation}
On the complete graph, the system can be fully described by $N_1(t),\,N_2(t),\,N_3(t)$, which, respectively, represent the number of agents with opinions $1,\,2,\,3$. 
Because the total number of agents $N$ is conserved:
\begin{equation}
N_1(t)+N_2(t)+N_3(t)=N, 
\label{eqn:number_def}
\end{equation}
we can alternatively introduce the normalized variables $c_1(t), \, c_2(t), \, c_3(t)$ defined as:
\begin{equation}
    c_i(t) = \frac{N_i(t)}{N},
\label{eqn:concentrations_def}
\end{equation}
which we call concentrations of opinions. From Eq. \eqref{eqn:number_def}:
\begin{equation}
\label{eqn:N_conservation}
    c_1(t) + c_2(t) + c_3(t) = 1,
\end{equation}
which reduces the number of independent variables to two. We choose $c_1(t)$ and $c_3(t)$ as independent variables due to the symmetry between the boundary states $1$ and $3$. 

To derive the temporal and stationary behavior of $c_1(t)$ and $c_3(t)$, we start, as usual, with the transition probabilities $\gamma_{i \rightarrow j}$, which represent the probability that a single agent will change their opinion from $i$ to $j$ \cite{Jed:Szn:19}. Analogously to $i$, variable $j\in \{1,2,3\}$.
Within the considered model, an agent can change their opinion only by one, hence to describe the change of concentration $c_i$ in an elementary time step, we only need to consider the transitions between $i$ and its adjacent states. In the limit of $N \rightarrow \infty$ the time derivatives of independent variables $c_1(t)$ and $c_3(t)$ can therefore be expressed in terms of only two transition probabilities, each constituting a following system of differential equations:
 \begin{equation}
     \left\{\begin{aligned}
         &\frac{dc_1}{dt} = \gamma_{2 \rightarrow 1} - \gamma_{1 \rightarrow 2}, \\
         &\frac{dc_3}{dt} = \gamma_{2 \rightarrow 3} - \gamma_{3 \rightarrow 2}.
         \label{eq:dc}
     \end{aligned}\right.
 \end{equation}

In general, Eq.\eqref{eq:dc} cannot be solved analytically.  Therefore we use a numerical approach to obtain trajectories, as well as stationary states:
\begin{equation}
\label{eq:vanihsing_derivatives}
     \left\{\begin{aligned}
         &\frac{dc_1}{dt} = 0, \\
         &\frac{dc_3}{dt} = 0 .
     \end{aligned}\right.
\end{equation}
The solutions of Eq. \eqref{eq:vanihsing_derivatives} correspond to fixed points that can be stable or unstable. The stability of fixed points will be also determined numerically by analyzing the eigenvalues of the corresponding Jacobian matrix. The analysis of fixed points and their stability was partially conducted using the DynamicalSystems.jl Julia software library \cite{Dat:18}. The explicit formulas for the transition probabilities $\gamma_{i \rightarrow j}$ depend on all parameters of the model ( $p,q,R_a,R_c$) and on the state of the system ($c_1,c_2,c_3$). For the infinitely large system within MFA, the probability of drawing $q$ voters with opinion $i$ is equal to $c_i^q$ and therefore:
\begin{align}
        \gamma_{2 \rightarrow 1} & = p\, c_2 \left[\frac{1}{2}c_2^q + \Theta(R_a - 1)c_1^q\right] + (1-p)\,c_2\Theta(R_c - 1)c_1^q,\label{eq:gamma_21}\\
        \gamma_{2 \rightarrow 3} & = p\, c_2 \left[\frac{1}{2}c_2^q + \Theta(R_a - 1)c_3^q\right] + (1-p)\,c_2\Theta(R_c - 1)c_3^q,\label{eq:gamma_23}\\
        \gamma_{1 \rightarrow 2} & = p\,c_1^{q+1} + (1-p)c_1\,\left[\Theta(R_c - 1)c_2^q + \Theta(R_c - 2) c_3^q\right],\label{eq:gamma_12}\\
        \gamma_{3 \rightarrow 2} & = p\,c_3^{q+1} + (1-p)\,c_3\left[\Theta(R_c - 1)c_2^q + \Theta(R_c - 2) c_1^q\right],\label{eq:gamma_32}
\end{align}
where $\Theta(x)$ denotes the Heaviside step function, which takes the value $0$ for $x<0$ and $1$ for $x \geq 0$.

\section{Results}
\label{sec:results}
\begin{figure}
  \includegraphics[width=\linewidth]{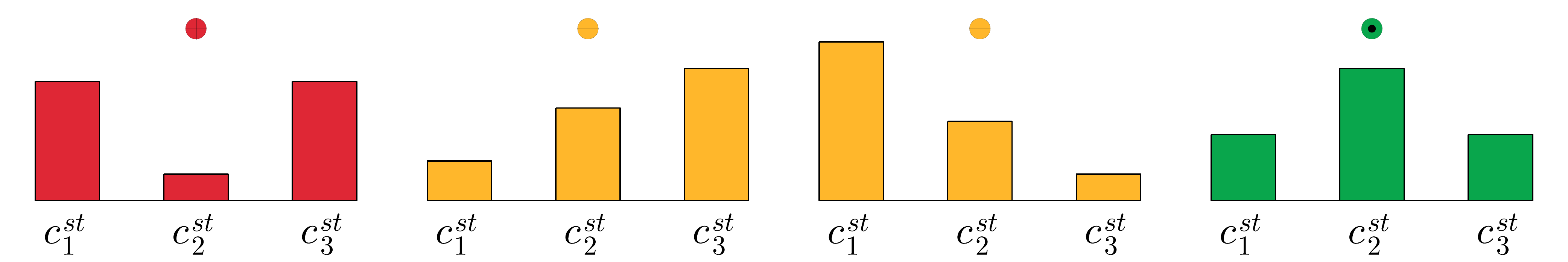}
  \caption{Schematic representation of three qualitatively different stationary opinion distributions: ($\oplus$, the leftmost panel) \textbf{P}olarization (\textbf{P}), ($\ominus$, two middle panels) \textbf{E}xtreme \textbf{D}ominance (\textbf{ED}), ($\odot$, the rightmost panel) \textbf{C}entral \textbf{D}ominance (\textbf{CD}).}
  \label{fig:op_dist}
\end{figure}

It appears that, in general, the system can evolve toward one of three different steady distributions of opinions, as shown in Fig. \ref{fig:op_dist}, for which we introduce the following names and abbreviations:
\begin{itemize}
\item
\textbf{P}olarization (\textbf{P}): $c_1=c_3>c_2$, the stable fixed point related to this structure is represented by $\oplus$
\item \textbf{E}xtreme \textbf{D}ominance (\textbf{ED}): $c_1>c_2>c_3$ or $c_3>c_2>c_1$, the stable fixed point related to this structure is represented by $\ominus$
\item \textbf{C}entral \textbf{D}ominance (\textbf{CD}): $c_2 > c_1=c_3$, the stable fixed point related to this structure is represented by $\odot$
\end{itemize}

A disordered state $c_1=c_2=c_3$ is not listed above, because it occurs that for a finite $q$ and $p \in (0, 1)$, such a state is a solution to Eq. \eqref{eq:vanihsing_derivatives} if and only if $R_a = 1, R_c = 2$ and $p = \frac{2}{3}$. This case is further discussed in Section \ref{ss:PD}.

Detailed results on possible steady states strongly depend on the ranges of bounded confidence $R_a$ and $R_c$, as will be shown in the following. For the three-state model $R_a$ and $R_c$ takes only three values, $0,1,2$, so in theory we have $9$ versions of the model. However, we can reduce this number because the model for $R_a=2$ is equivalent to the model with $R_a=1$, as described in the caption of Tab. \ref{tab:tran_anti}. Furthermore, for $R_c=0$ there are no transitions caused by conformity, as indicated by Tab. \ref{tab:tran_con}. Therefore, in the following part of the work we will analyze  four cases: (1) $R_a=0, R_c=1$, (2) $R_a=R_c=1$, (3) $R_a=0, R_c=2$ and (4) $R_a=1, R_c=2$.

As stated previously, homophily is the tendency to conform to people that are already similar to us. The model we propose entails homophily for $R_c = 1$, while for $R_c = 2$ it is no longer the case, as the target conforms to the source despite the similarity. Hence two versions of the model for which $R_c=1$ are the most interesting from a social point of view and we will therefore present the results for these versions in more detail and also verify the MFA results through Monte Carlo simulations.

\subsection{Phase portraits for $R_a=0, R_c=1$}
\begin{figure}
    \centering
    \includegraphics[width = 0.8\linewidth]{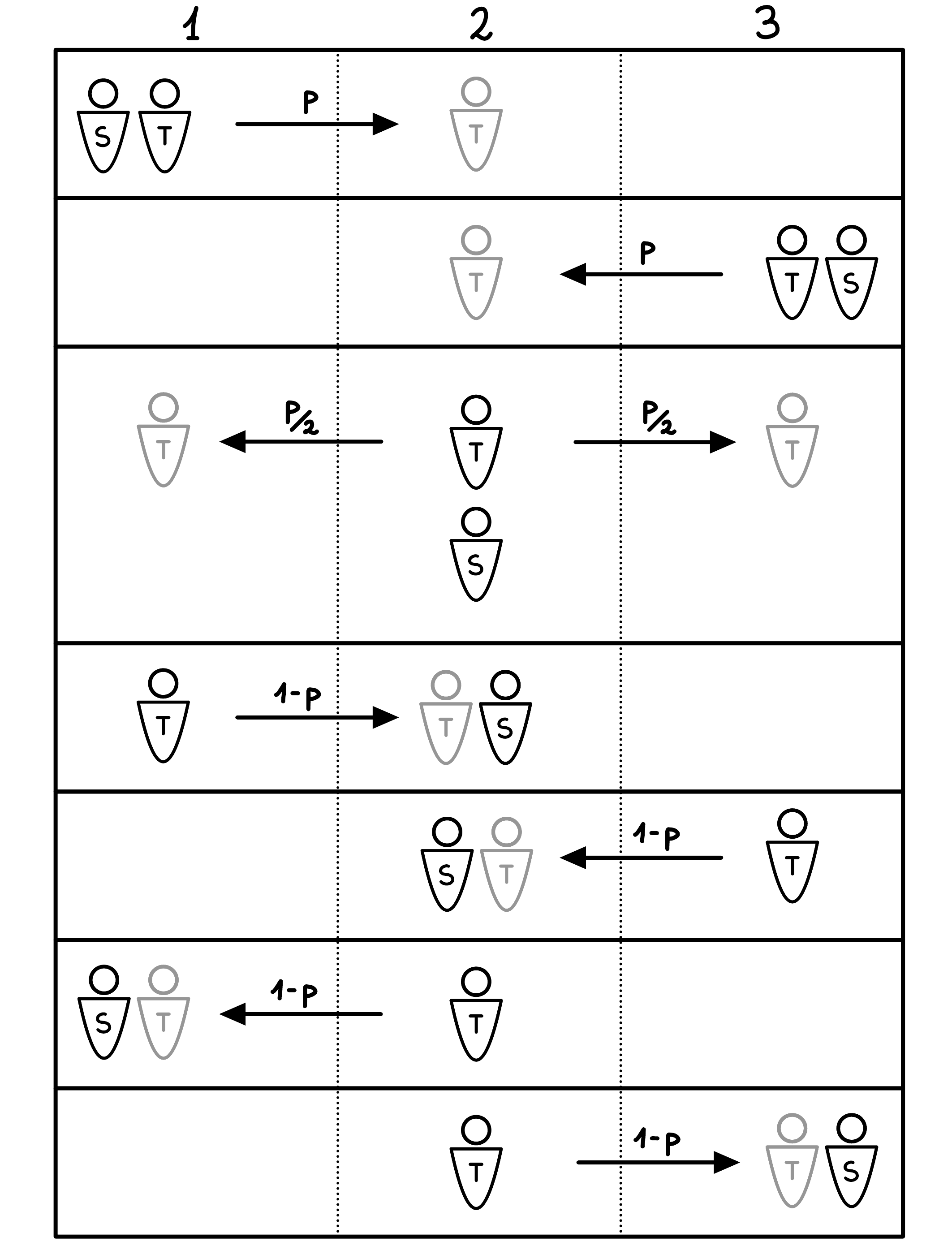}
    \caption{All possible changes for $R_a = 0,\, R_c = 1$, where T denotes the target and S the source of influence ($q$-panel). Panels are partitioned into three areas, marked by $1,\,2,\,3$, denoting opinion states. Annotations above arrows denote the probability that the corresponding opinion change occurs, given the opinions of the target and the source.}
    \label{fig:scheme_ra0}
\end{figure}

In this case, all possible transitions are presented in Fig. \ref{fig:scheme_ra0} and Eq. \eqref{eq:dc} takes form:
\begin{equation}
\label{eqn:ra0_time_evo}
     \left\{\begin{aligned}
         \frac{dc_1}{dt} = p &\left(\frac{1}{2}c_2^{q+1} - c_1^{q+1}\right) + \left(1-p\right)\left[c_2c_1^q - c_1c_2^q\right], \\
         \frac{dc_3}{dt} = p &\left(\frac{1}{2}c_2^{q+1} - c_3^{q+1}\right) + \left(1-p\right)\left[c_2c_3^q - c_3c_2^q\right],
     \end{aligned}\right.
\end{equation}
where $c_2 = 1 - c_1 - c_3$. 

\begin{figure}
\centering
\includegraphics[width = \textwidth]{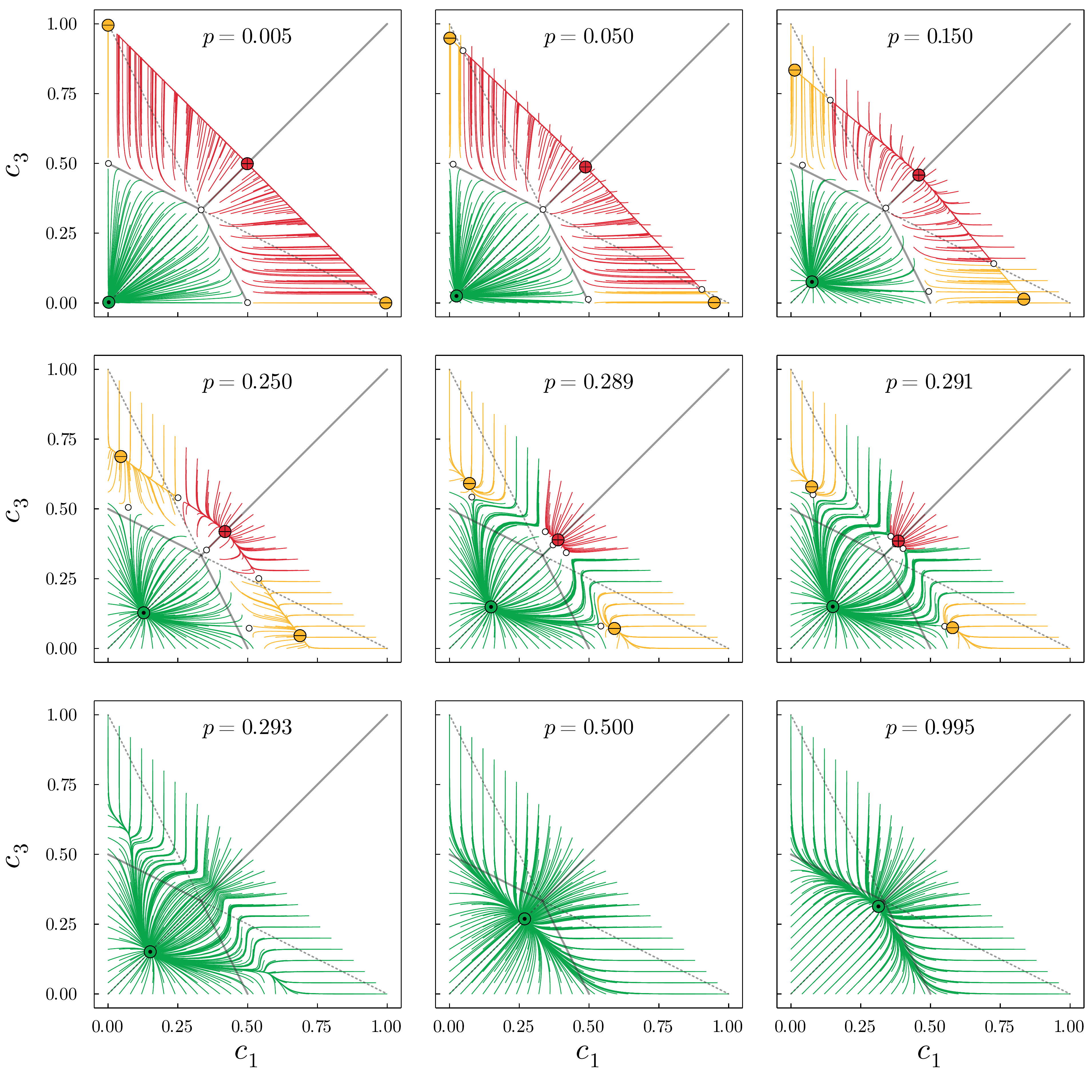}
\caption{Phase portraits for $R_a = 0, R_c = 1$ and $q = 3$. Each panel corresponds to different value of the probability of anticonformity $p$, as indicated inside of each panel. Solid colored lines represent the trajectories obtained from Eq. \eqref{eqn:ra0_time_evo}, and the colored circles indicate the stable fixed points. The colors and symbols inside the circles representing stable fixed points correspond to different stationary opinion distributions, as presented in Fig. \ref{fig:op_dist}. Empty circles indicate unstable fixed points. The critical point, above which there is only one stable fixed point $\odot$, corresponding to \textbf{C}entral \textbf{D}ominance, $p^* = p^*(q = 3, R_a = 0, R_c = 1) \in (0.291,0.293)$.}
\label{fig:panels_ra0}
\end{figure}

The corresponding phase portraits, showing the trajectories for various initial conditions, as well as stable and unstable fixed points, are presented in Fig. \ref{fig:panels_ra0}. In this figure we present results for the size of the influence group $q=3$, but qualitatively the same can be obtained for other values of $q$, which will be discussed in Section \ref{ss:PD}. We see that there is a critical point $p=p^*$, at which three bifurcations simultaneously occur: two stable fixed points $\ominus$, related to \textbf{E}xtreme \textbf{D}ominance and a fixed point $\oplus$, related to \textbf{P}olarization, annihilate with corresponding unstable  points. Unfortunately, we were not able to calculate the critical point analytically, and therefore the precise value of $p^*$ cannot be provided. However, the upper and lower bound of $p^*$ can be obtained by the stability analysis of fixed points. As we see in Fig. \ref{fig:panels_ra0}, for $q=3$ the critical point satisfies $0.291 < p^* < 0.293$. For $p>p^*$, there is only one stable fixed point $\odot$, corresponding to \textbf{C}entral \textbf{D}ominance. For $p<p^*$, there are four basins of attraction, each corresponding to qualitatively different distributions of opinions, shown schematically in Fig. \ref{fig:op_dist}. For small values of $p$ there is a large basin of attraction corresponding to \textbf{P}olarization and simultaneously the population of voters with central opinion is small ($c_1=c_3 \rightarrow 1/2$). 
With increasing probability of anticonformity $p$ the fixed point $\oplus$,  moves away from the strong polarization point ($c_1=c_3\rightarrow 1/2$) and simultaneously its basin of attraction shrinks. This leads to the conclusion that conformity promotes polarization.

\subsection{Phase portraits for $R_a = R_c = 1$}
\begin{figure}
    \centering
    \includegraphics[width = 0.8\linewidth]{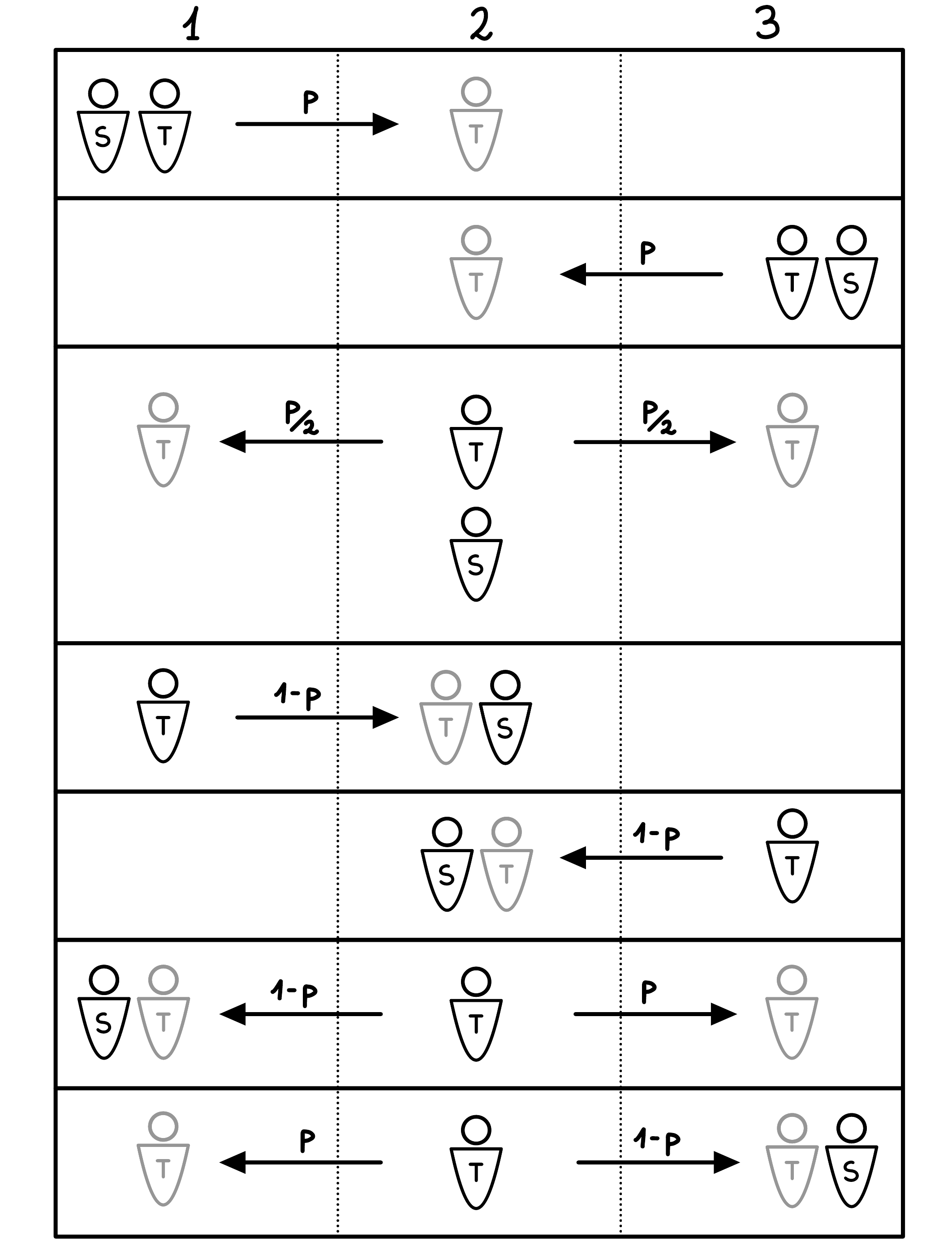}
    \caption{All possible changes for $R_a =  R_c = 1$, where T denotes the target and S the source of influence ($q$-panel). Panels are partitioned into three areas, marked by $1,\,2,\,3$, denoting opinion states. Annotations above arrows denote the probability that the corresponding opinion change occurs, given the opinions of the target and the source.}
    \label{fig:scheme_ra1}
\end{figure}

\begin{figure}
    \centering
    \includegraphics[width = \textwidth]{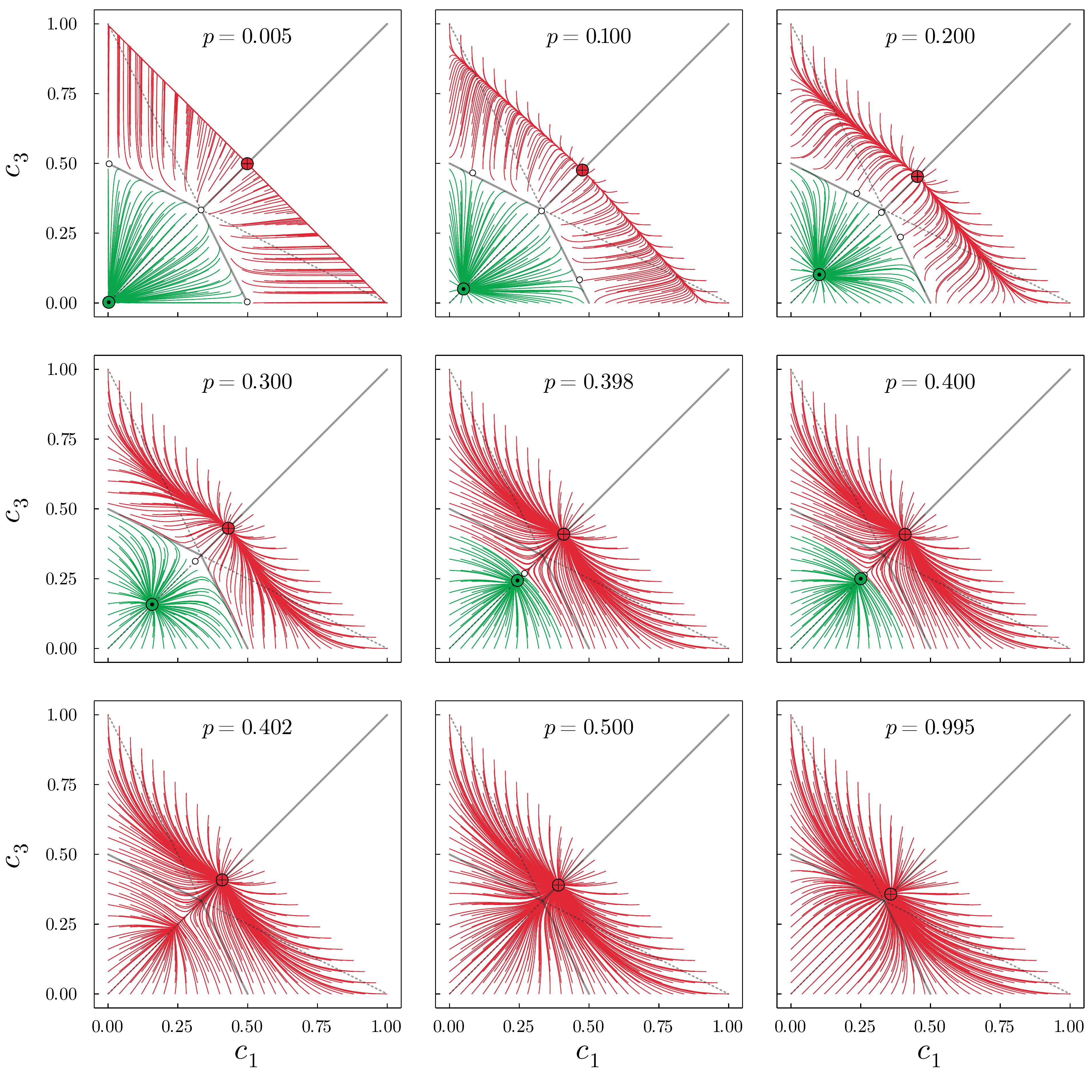}
    \caption{Phase portraits for $R_a = R_c = 1$ and $q = 3$. Each panel corresponds to different value of the probability of anticonformity $p$, as indicated inside of each panel. Solid colored lines represent the trajectories obtained from Eq. \eqref{eqn:ra1_time_evo}, and the colored circles indicate the stable fixed points. The colors and symbols inside the circles representing stable fixed points correspond to different stationary opinion distributions, as presented in Fig. \ref{fig:op_dist}. Empty circles indicate unstable fixed points. The critical point, above which there is only one stable fixed point  $\oplus$, related to \textbf{P}olarization, $p^* = p^*(q = 3, R_a = 1, R_c = 1) \in (0.400,0.402)$.}
    \label{fig:panels_ra1}
\end{figure}

In this case, all possible transitions are presented in Fig. \ref{fig:scheme_ra1} and Eq. \eqref{eq:dc} takes form:
\begin{equation}
\label{eqn:ra1_time_evo}
     \left\{\begin{aligned}
         \frac{dc_1}{dt} = p &\left(c_2c_3^q + \frac{1}{2}c_2^{q+1} - c_1^{q+1}\right) + \left(1-p\right)\left[c_2c_1^q - c_1c_2^q\right], \\
         \frac{dc_3}{dt} =  p &\left(c_2c_1^q + \frac{1}{2}c_2^{q+1} - c_3^{q+1}\right) + \left(1-p\right)\left[c_2c_3^q - c_3c_2^q\right],
     \end{aligned}\right.
\end{equation}
where, as always, $c_2 = 1 - c_1 - c_3$. The corresponding phase portraits, showing trajectories for various initial conditions, as well as stable and unstable fixed points, are presented in Fig. \ref{fig:panels_ra1}. Increasing the range $R_a$ for anticonformity from $R_a = 0$ to $R_a = 1$ qualitatively influences the results. In Fig.~\ref{fig:panels_ra1} it can be seen that there are now only two attractors for $p < p^*$, corresponding to \textbf{P}olarization and \textbf{C}entral \textbf{D}ominance. It means that extreme opinion cannot dominate in the stationary state, independently of the initial conditions. A similar result was obtained for the three-state $q$ voter model with independence and bounded confidence \cite{Don:Lip:Szn:22}. Contrary to the behavior observed for $R_a = 0$, the basin of attraction for \textbf{P}olarization increases with $p$ and for $p>p^*(q)$ \textbf{P}olarization is the only stationary solution. However, similarly to $R_a=0$ strong polarization, understood as $c_1=c_3 \rightarrow 1/2$, occurs only for small values of $p$.

\subsection{Phase portraits for $R_c=2$}
\begin{figure}
    \centering
    \includegraphics[width = \linewidth]{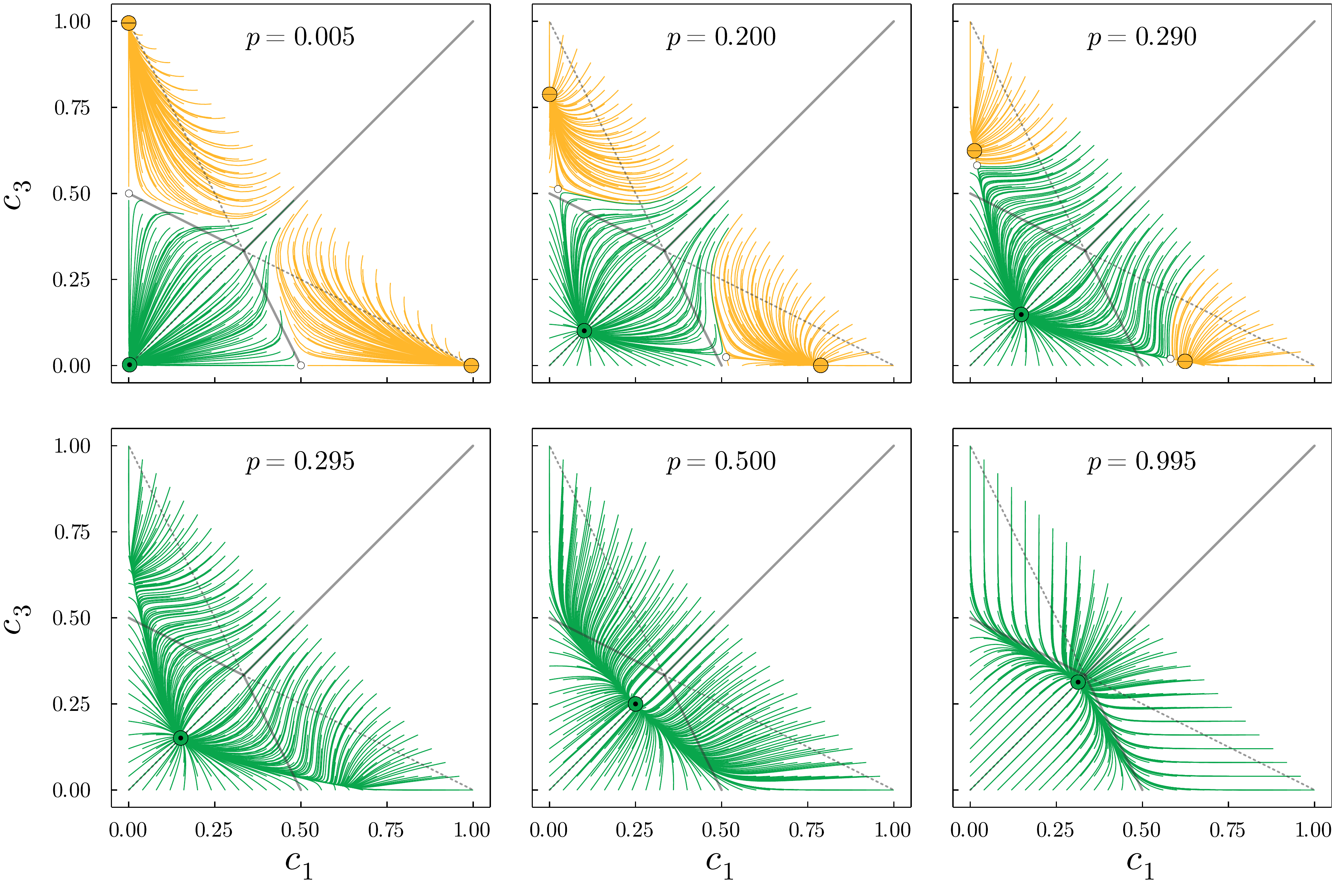}
    \caption{Phase portraits for $R_a = 0,\,R_c = 2$ and $q = 3$. Each panel corresponds to different value of the probability of anticonformity $p$, as indicated inside of each panel. Solid colored lines represent the trajectories and the colored circles indicate the stable fixed points. The colors and symbols inside the circles representing stable fixed points correspond to different stationary opinion distributions, as presented in Fig. \ref{fig:op_dist}. Empty circles indicate unstable fixed points. The critical point, above which there is only one stable fixed point $\odot$, corresponding to \textbf{C}entral \textbf{D}ominance, $p^* = p^*(q = 3, R_a = 0, R_c = 2) \in (0.290,0.295)$.}
    \label{fig:panels_rc2_ra0}
\end{figure}

\begin{figure}
    \centering
    \includegraphics[width = \linewidth]{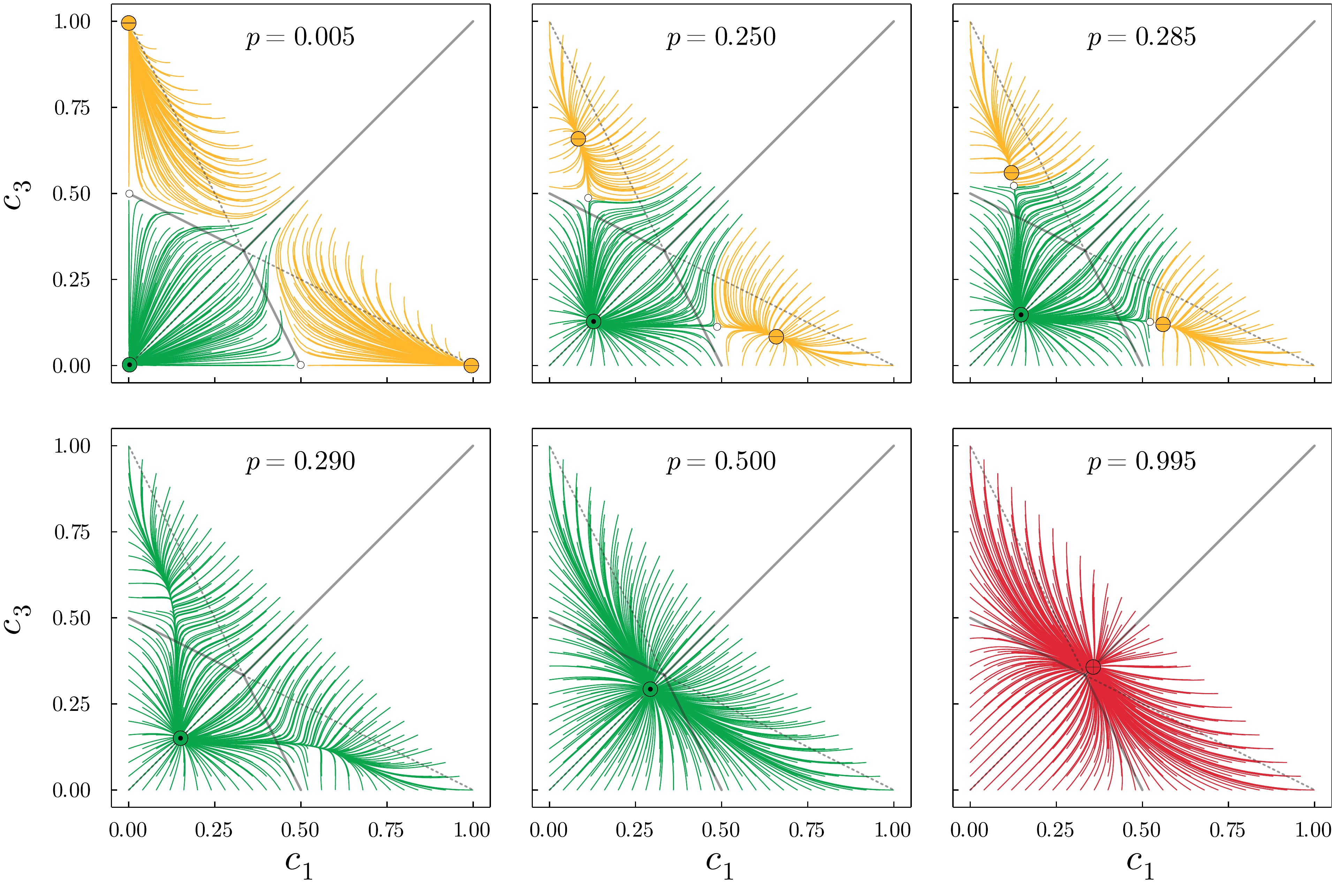}
    \caption{Phase portraits for $R_a = 1,\,R_c = 2$ and $q = 3$. Each panel corresponds to different value of the probability of anticonformity $p$, as indicated inside of each panel. Solid colored lines represent the trajectories and the colored circles indicate the stable fixed points. The colors and symbols inside the circles representing stable fixed points correspond to different stationary opinion distributions, as presented in Fig. \ref{fig:op_dist}. Empty circles indicate unstable fixed points. The critical point, above which there is only one stable fixed point  $p^* = p^*(q = 3, R_a = 1, R_c = 2) \in (0.285,0.290)$.}
    \label{fig:panels_rc2_ra1}
\end{figure}

The case of $R_c=2$ is less compatible with the idea of homophily than the case of $R_c=1$ and thus less interesting from the social point of view. However, for the sake of completeness, we will also briefly present the results for this case. We can carry out the entire analysis in exactly the same way as for $R_c=1$, but we focus here only on the  phase portraits, which are presented in Fig. \ref{fig:panels_rc2_ra0} for $R_a=0$ and  in Fig. \ref{fig:panels_rc2_ra1} for $R_a=1$. For both values of $R_a$ we obtain the same bifurcation at $p=p^*$: for $p<p^*$ \textbf{E}xtreme \textbf{D}ominance or \textbf{C}entral \textbf{D}ominance is reached depending on the initial conditions. The only difference between $R_a=0$ and $R_a=1$ appears for $p>2/3>p^*$: for $R_a=0$ \textbf{C}entral \textbf{D}ominance is reached for any $p > p^*$, while for $R_a=1$ the stationary concentration of the central opinion decays with $p$ and eventually \textbf{C}entral \textbf{D}ominance smoothly (without bifurcation) transforms into \textbf{P}olarization. However, in principle, the social structure does not change much because this polarization is very weak: $c_1=c_3$ is only slightly larger than $c_2$, as shown in the last panel of Fig. \ref{fig:panels_rc2_ra1}. 


\subsection{Monte Carlo results}
In this section we will compare the results obtained within two independent methods: numerical solutions of the MFA equations \eqref{eq:dc} and \eqref{eq:vanihsing_derivatives}, as well as Monte Carlo (MC) simulations. In the latter case, we will present results for $N=10^5$, because then the MC results overlap the MFA results that correspond to $N \rightarrow \infty$, similarly to other papers on the $q$-voter model \cite{Now:Sto:Szn:21,Now:Szn:22}. 

\begin{figure}[h!]
    \centering
    \includegraphics[width = \linewidth]{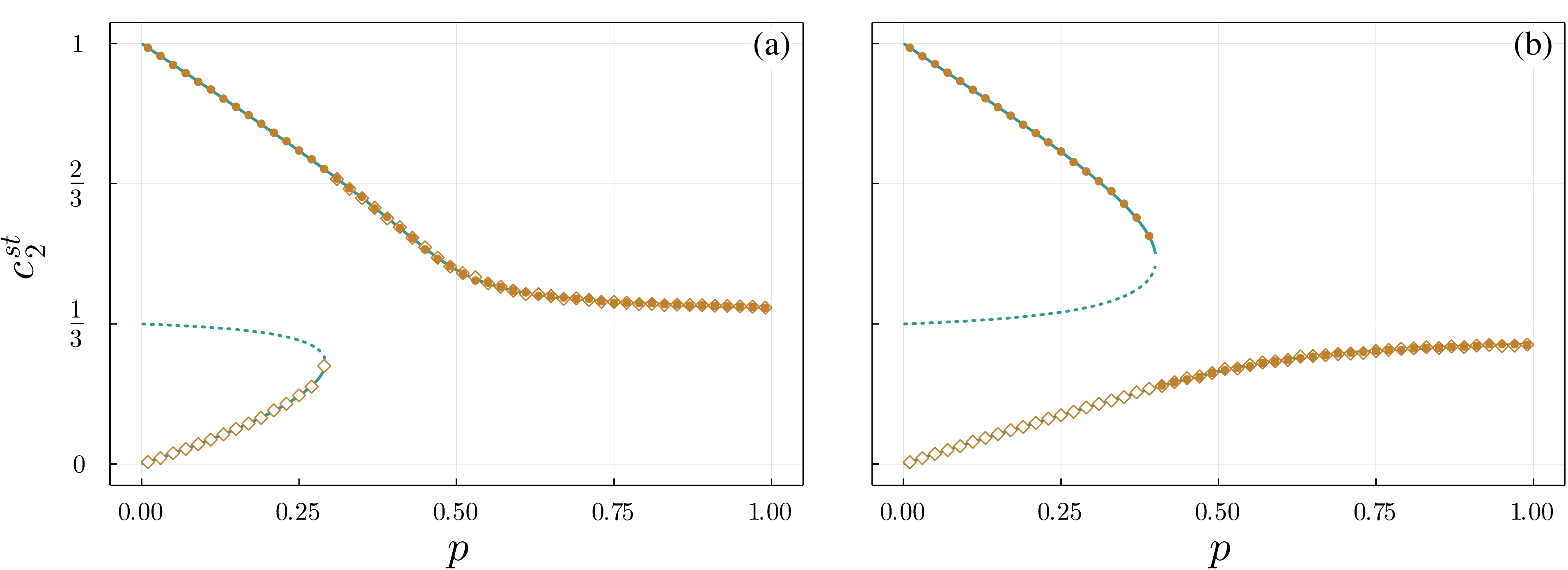}
    \caption{Stationary value of the central opinion concentration $c_2^{st}$ as a function of the probability of anticonformity $p$, satisfying the condition $c_1^{st} = c_3^{st}$ for $q=3,\,R_c=1$ and (a) $R_a=0$, (b) $R_a=1$. MFA results are represented by solid (stable fixed points) and dotted (unstable fixed points) lines. Symbols represent the MC results from a single run for the system size $N = 10^5$ and equilibration time $\tau = 10^5$ MCS. Results for two different initial conditions are shown: full circles correspond to  $c_1(0) = c_3(0) = 0$, and empty diamonds to $c_1(0) = c_3(0) = 0.5$.}
    \label{fig:q3_mc}
\end{figure}

In order to limit the parameter space, for which MC simulations are performed, we will present results satisfying $c_1^{st} = c_3^{st}$, i.e., stationary concentrations corresponding to \textbf{CD} or \textbf{P}. Employing this symmetry, which is seen from the description of the model, as well as from phase portraits, allows us to present the fixed points of the system only in terms of $c_2$, as proposed in \cite{Don:Lip:Szn:22}. In Fig.~\ref{fig:q3_mc} the dependence between the stationary value of the central opinion concentration $c^{st}_2$ and the probability of anticonformity $p$ is shown. The fixed points obtained within MFA are complemented with the results of MC simulations to confirm the agreement between the methods.  

From a bifurcation point of view, the results for the two versions of the model, $R_a=0$ and $R_a=1$ are different and results almost look like mirror images relative to the $c^{st}_2=1/3$. If we look at the actual values of $c^{st}_2$ for a given $p$ there are not so much differences for $R_a=0$ and $R_a=1$. For $p<p^*$ there are two stable values of $c^{st}_2$, which approach $0$ (maximal \textbf{P}olarization) and $1$ (maximal \textbf{C}entral \textbf{D}ominance) for decreasing values of $p$. For $p>p^*$ the system never reaches the state of disagreement $c_1=c_2=c_3=1/3$, which is a typical behavior of the three-state $q$-voter model with bounded confidence and independence. However, for both values of $R_a$ the system is approaching this state with the increasing value of $p$: for $R_a=0$ from above $c^{st}_2=1/3$ and for $R_a=1$ from below $c^{st}_2=1/3$. Formally, this corresponds to two different social structures: \textbf{C}entral \textbf{D}ominance for $R_a=0$ and \textbf{P}olarization for $R_a=1$ but the differences between $c_1,\,c_2,\,c_3$ are very small.

\subsection{Phase diagrams} \label{ss:PD}
Because MFA and MC results clearly overlap, as shown in Fig. \ref{fig:q3_mc}, in this section we will use MFA to compare all four versions of the model and evaluate the role of the size of the influence group $q$, which was neglected so far. Moreover, we will present phase diagrams based on MFA for all four versions of the model.

\begin{figure}
    \centering
    \includegraphics[width = \linewidth]{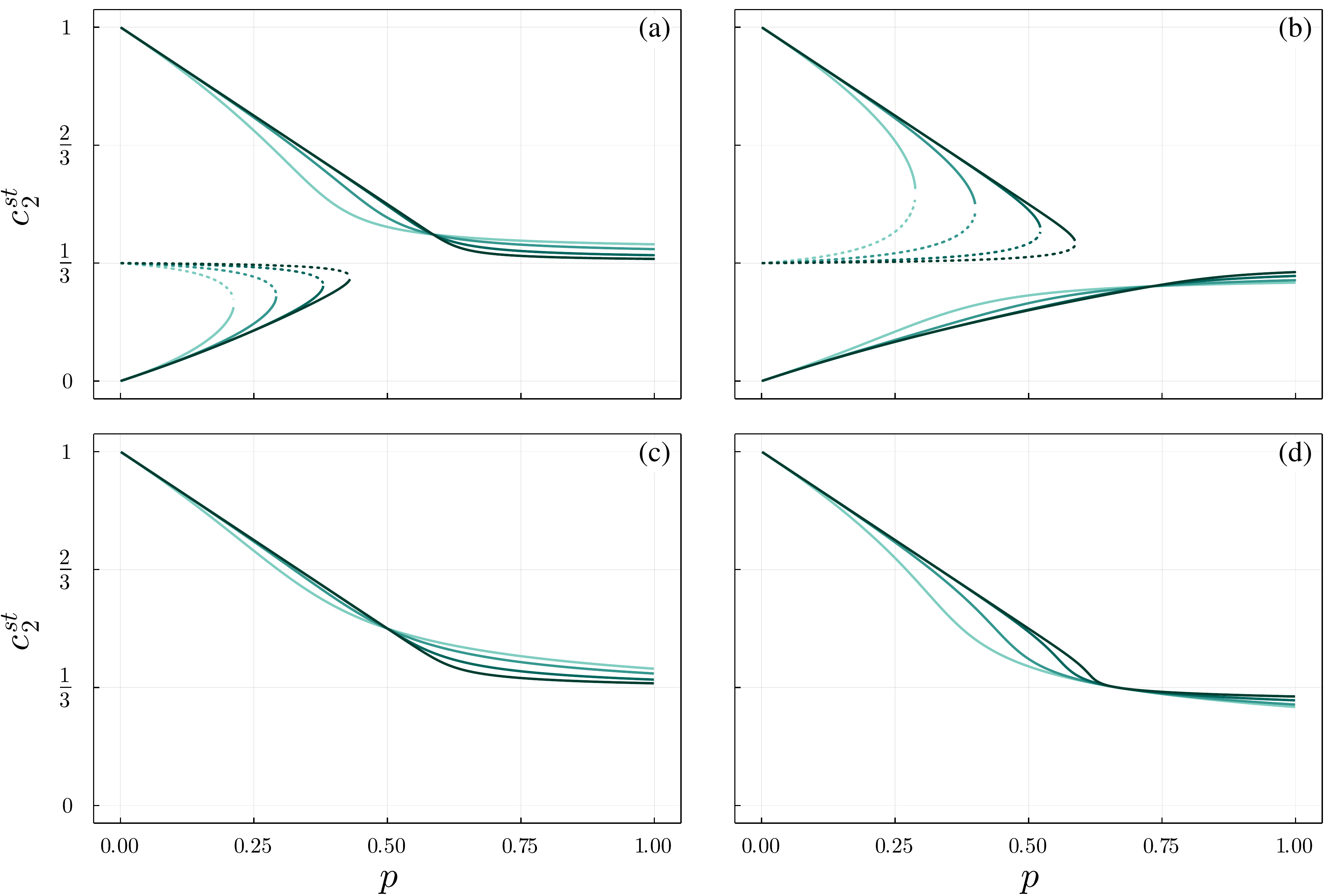}
    \caption{Stationary value of the central opinion concentration $c_2^{st}$ as a function of the probability of anticonformity $p$, satisfying the condition $c_1^{st} = c_3^{st}$ for (a) $R_a = 0, \, R_c = 1$, (b) $R_a = 1, \, R_c = 1$, (c) $R_a = 0, \, R_c = 2$, (d) $R_a = 1, \, R_c = 2$. Solid lines represent stable fixed points, whereas dotted lines unstable fixed points. Results for $q = 2,\,3,\,6,\,12$ are plotted with lines, which intensity increases with $q$.}
    \label{fig:cp_mfa}
\end{figure}

We start with presenting in Fig. \ref{fig:cp_mfa} the stationary value of the central opinion concentration $c_2^{st}$ as a function of the probability of anticonformity $p$, satisfying the condition $c_1^{st} = c_3^{st}$ for different values of $q$. We see that if we focus on the region of phase space corresponding to the line $c_1 = c_3$, the size of the influence $q$ seems to have much larger impact for the models with $R_c=1$. In this case, independently on $R_a$, the critical point $p^*=p^*(q)$ is the increasing function of $q$, similarly as in other versions of the $q$-voter model with anticonformity but without bounded confidence \cite{Now:Szn:22}. For $R_c=2$ no critical points are visible and $q$
virtually does not affect the results.

However, if we look at phase portraits for the entire space $\left(c_1,c_3\right)$, as shown in Figs. \ref{fig:panels_ra0}, \ref{fig:panels_ra1}, \ref{fig:panels_rc2_ra0}, \ref{fig:panels_rc2_ra1}, and not just the line $c_1 = c_3$, we see much richer behavior. In particular, for both cases with $R_c=2$ there is a bifurcation at $p<p^*(q)$ (the value of $p^*$ is different for $R_a = 0$ and $R_a = 1$), during which \textbf{E}xtreme \textbf{D}ominance vanishes. 

\begin{figure}
    \centering
    \includegraphics[width = \linewidth]{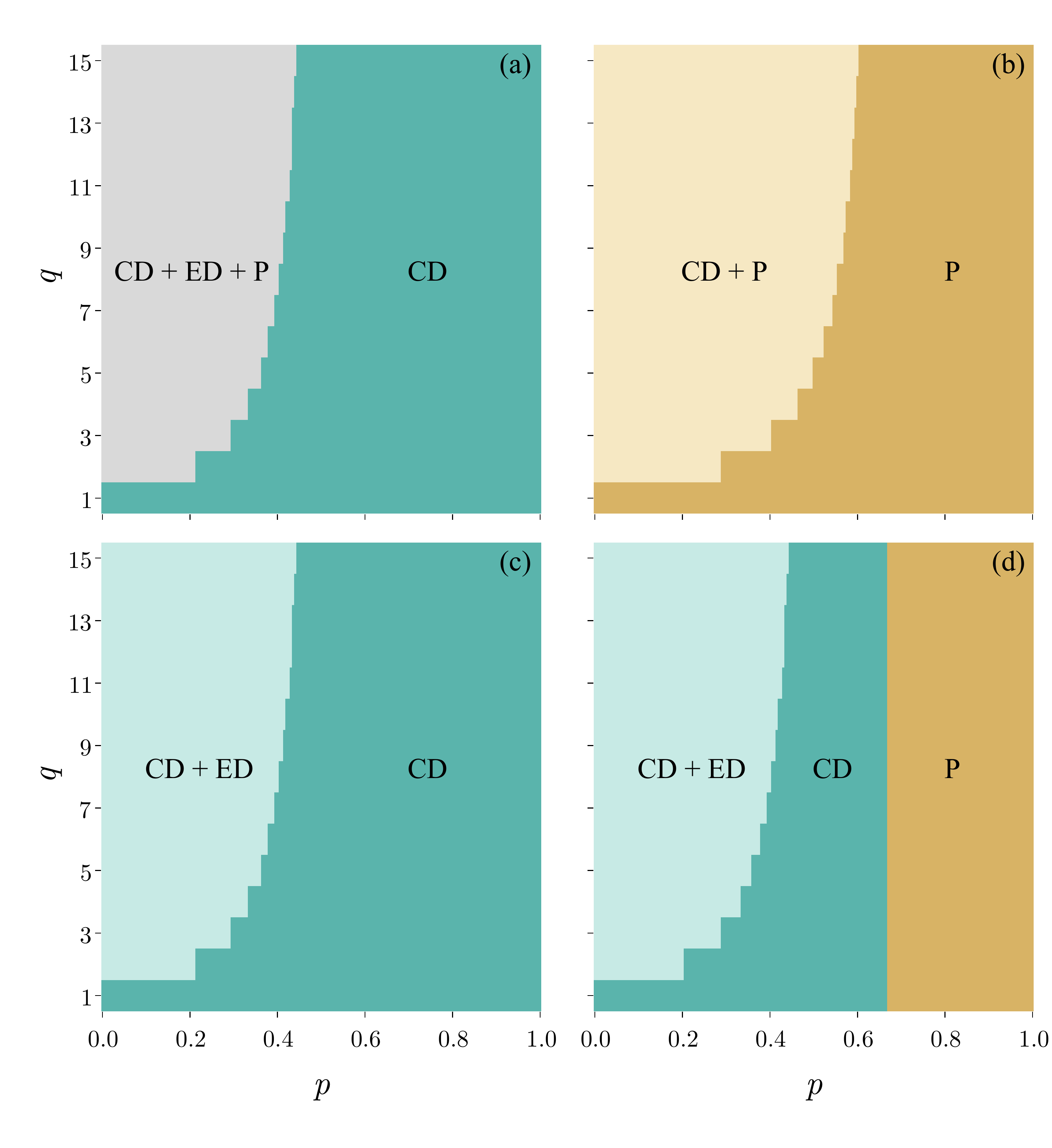}
    \caption{Phase diagrams showing the coexistence of different stable fixed points that occur for given values of model parameters $p, q$ and: (a) $R_a = 0, \, R_c = 1$, (b) $R_a = 1, \, R_c = 1$, (c) $R_a = 0, \, R_c = 2$, (d) $R_a = 1, \, R_c = 2$. The phases are denoted by colors and labeled with the abbreviations introduced at the beginning of Sec. \ref{sec:results} and repeated in the caption of Fig. \ref{fig:op_dist}.}
    \label{fig:PD}
\end{figure}

Therefore, to compare models systematically, we decided to prepare phase diagrams, as shown in Fig. \ref{fig:PD}. To derive them, we performed a similar analysis to the one presented in Figs. \ref{fig:panels_ra0},\ref{fig:panels_ra1},\ref{fig:panels_rc2_ra0} and \ref{fig:panels_rc2_ra1}. To clarify this, let us focus for a while on Fig. \ref{fig:panels_ra0}, which corresponds to $R_a=0,R_c=1$ and $q=3$. We see that for $p<p^*$, where $p^* \in (0.291,0.293)$ there are three types of stable fixed points: \textbf{C}entral \textbf{D}ominance, \textbf{E}xtreme \textbf{D}ominance and \textbf{P}olarization. Therefore, we denote this phase as \textbf{CD+ED+P}. For $p>p^*$ there is only one fixed point, \textbf{C}entral \textbf{D}ominance, and thus we denote this phase \textbf{CD}. Conducting a stability analysis of fixed points for different values of $q$ we can obtain the critical value $p^*=p^*(q)$, which gives us the border between the \textbf{CD+ED+P} and \textbf{CD} phases, as shown in the upper left panel of Fig. \ref{fig:PD}. In the same way, we construct phase diagrams for other values of $R_a,R_c$.

As seen in Fig. \ref{fig:PD}, $p^*=p^*(q)$ is an increasing function of $q$. In general, in all four models, there is only one critical point $p^*=p^*(q)$, which is always an increasing function of $q$. The difference between models is that at this critical point separates different phases, as shown in Fig. \ref{fig:PD}, and is related to different numbers of bifurcations. 

In panel (d) of Fig. \ref{fig:PD}, that is, for $R_a=1$ and $R_c=2$, we see three, not two phases, which might suggest an additional critical point. However, as seen in Fig. \ref{fig:panels_rc2_ra1}, this is not the case -- there is no bifurcation responsible for the appearance of this phase. It appears because the stable fixed point related to \textbf{CD} moves continuously along the line $c_1= c_3$, and at $p=p_{eq}=2/3$ (independently on $q$) crosses the point of equality $c_1 = c_2 = c_3$. The specific value of $p_{eq}=2/3$ stems directly from the construction of the model. When the range of bounded confidence spans the entire opinion space ($R_a=1,\,R_c=2$) then all transitions are possible and the fixed point in the phase space region $c_1 = c_3$ corresponds to a solution of the following equation:
\begin{equation}
\label{eq:ra1_rc2_fp}
    \left(c^{st}_2 - c^{st}_1\right){c^{st}_1}^q + \left(\frac{p}{2} - c^{st}_1\right){c^{st}_2}^q = 0.
\end{equation} The former term on the left-hand side of Eq. \eqref{eq:ra1_rc2_fp} arises from the interactions with the source which holds an extreme opinion, whereas the latter with the middle one. Keeping in mind that the number of agents is conserved, which is described by Eq. \eqref{eqn:N_conservation}, one can directly check that $c^{st}_1$ and $c^{st}_2$ must be nonzero and since $c_1 = c_2 \iff c_2 = \frac{1}{3}$, the fixed point $c^{st}_2$ clearly must satisfy: $\left(p < \frac{2}{3} \;\wedge\; c^{st}_2 > \frac{1}{3}\right) \vee \left(p = \frac{2}{3} \;\wedge\; c^{st}_2 = \frac{1}{3}\right) \vee \left(p > \frac{2}{3} \;\wedge\; c^{st}_2 < \frac{1}{3}\right)$. Otherwise, the terms in Eq. \eqref{eq:ra1_rc2_fp} do not cancel out and it cannot equate to zero.

\FloatBarrier
\section{Discussion}
In this work we asked the question about the role of anticonformity in shaping social polarization in the discrete-choice scenario.  M\"as and Flache demonstrated within an agent-base model and a group discussion experiment that polarization can be explained without negative influence \cite{Mas:Fla:13}. Similar result has been obtained in several models of continuous opinion dynamics with bounded confidence \cite{Def:etal:02, Heg:Kra:02, Fla:etal:17}. Within continuous opinion models, even if random deviations from bounded confidence are taken into account, it is still possible to explain polarization without negative influence if only some noise is added to the system \cite{Kur:Mas:Lor:16}. In fact, it is possible to explain polarization even completely disregarding bounded confidence and negative interactions \cite{Ban:Olb:19}. However, in this paper we do not deal with continuous opinions. Instead we focus on the three-state variables, which can describe the choice between three ranked alternatives: two extremes and the central (moderate) one.
Our model is not the first model with three-state opinions and bounded confidence -- several others have already been proposed \cite{Don:Lip:Szn:22, Vaz:Kra:Red:03, Vaz:Red:04, Mob:11, Cro:14, Bal:Pin:Sem:15, Rad:Don:21}.Below we shortly review them and explain their relation to our model.

\subsection{Related models}
Vazquez et al. \cite{Vaz:Kra:Red:03} introduced a three-state voter-like dynamics with conformity and bounded confidence on a grid. Within this model, a target agent adopts the opinion of their randomly selected neighbor if the difference between their opinions equals one. Such a simple rule leads either to ultimate consensus or to a frozen population of leftists and rightists, which can be interpreted as an extreme polarization. Neither ultimate consensus nor an extreme polarization is seen in real societies, and thus the model is certainly oversimplified. However, it provides a minimalist description for how consensus or polarization can be achieved. It was later examined by Vazquez and Redner \cite{Vaz:Red:04} on the complete graph, which corresponds to our model in the case of $p=0, q=1,  R_c = 1$. Note that the value of $R_a$ is not relevant since only conformity is allowed. Mobilia \cite{Mob:11} further generalized this model with the inclusion of bias towards polarization or centrism. In this paper, we do not introduce any kind of bias, but such an extension could be done in the future.

The most crucial difference between our model and the one proposed Vazquez et al. \cite{Vaz:Kra:Red:03} is the presence of not only conformity, but also anticonformity. The competition between these two social forces allows for more realistic social structures, under which all 3 opinions coexist, as shown in Fig. \ref{fig:op_dist}. However, the three-state opinion models in which both conformity and anticonformity are present, have also been studied previously \cite{Cro:14, Bal:Pin:Sem:15}. The main difference between them and the one introduced in this paper is the definition of the middle state. Unlike our approach, it is not interpreted as a distinct opinion, but rather as a lack thereof. In \cite{Cro:14} the undecided agent cannot influence others, while in \cite{Bal:Pin:Sem:15} they can be more easily persuaded than those with established opinions. The mechanisms of social response differ as well, in \cite{Bal:Pin:Sem:15} positive interactions occur exclusively between similar agents and negative interactions between dissimilar ones, while in \cite{Cro:14} the type of interaction between each pair of agents is given by a quenched random variable independent on their relative opinions. Hence, both \cite{Bal:Pin:Sem:15} and \cite{Cro:14} cannot be translated into our model. Contrary to our results, the coexistence of non-zero steady concentrations of all three opinions has not been reported in \cite{Cro:14, Bal:Pin:Sem:15}. Moreover, in both models, a source of influence consists of a single neighbor, as in \cite{Vaz:Kra:Red:03,Vaz:Red:04}, which corresponds to the special case of qVM with $q=1$. Therefore, the role of the size of the influence group could not be studied there.

Within three-state qVM with BC, only the interplay between conformity and independence (noise) has been examined so far \cite{Don:Lip:Szn:22, Rad:Don:21}. Although the model introduced in \cite{Don:Lip:Szn:22} can lead to opinion polarization, it does not exhibit transitions between polarization and central dominance. Furthermore, in contrast to the model studied here, in \cite{Don:Lip:Szn:22} polarization is never observed as the only stable fixed point whose basin of attraction extends over the entire space of initial conditions.

\subsection{Conclusions}
We believe that taking into account both types of social response (positive and negative) is important when modeling the dynamics of opinion within an organization, because disagreement can be fruitful and improve organization performance \cite{Lan:Fou:Hol:18,Min:Gin:22}. On the other hand, what is not desirable is polarization, which can hinder any common decision from being made. That is why we wanted to check how anticonformity influences polarization. Although a negative influence appearing with a certain probability $p$ has already been considered in some three-state models \cite{Cro:14}, to our knowledge independent ranges of conformity and anticonformity interactions are the novelty introduced in this work.

As a starting point for our model we have chosen the $q$-voter model, which seems to be particularly interesting from the psychological point of view, as reviewed in \cite{Jed:Szn:19}. Within the proposed model, we showed that strong polarization appears when two conditions are met: 
\begin{enumerate}
    \item The range of bounded confidence for conformity $R_c=1$, which corresponds to homophily.
    \item The probability of anticonformity $p$ is low. 
\end{enumerate}
Although in the case of $R_c = R_a = 1$ the opinion distribution always evolves towards \textbf{P}olarization for $p>p^*(q)$, the system is not considered to be strongly polarized, as $c_2^{st}$ is relatively high. If $R_a=0$, which means that anticonformity appears only if the source has the same opinion as the target, then an intermediate level of anticonformity $p$, which is right above the critical value $p^*(q)$, promotes domination of the moderate opinion, as shown in Fig. \ref{fig:panels_ra0}. In such a case, domination of the central opinion is not only strong but also reached from arbitrary initial conditions. It means that the occasional anticonformity among the organization members not only prevents opinion polarization from arising, but also promotes arriving at a general agreement -- a compromise is reached even in the case of initial total polarization. This is a particularly interesting result, because $R_a=0$ denotes the anticonformity that is caused by asserting moderate uniqueness. Interestingly, it was shown experimentally that people feel better when they see themselves as moderately unique \cite{Mar:Men:14}, indicating that $R_a=0$ may be psychologically justified. The results we present lead us to the conclusion that challenging the viewpoints of like-minded peers and rebelling against unanimous groups can prove to be valuable mechanisms in the process of collective opinion formation, provided that they are not too frequent.

\section*{Autorship contribution statement} \textbf{Arkadiusz Lipiecki:} Conceptualization, Methodology, Software, Visualization, Investigation, Writing – original draft. \textbf{Katarzyna Sznajd-Weron:} Supervision, Writing – review and editing.

\section*{Declaration of competing interests} Katarzyna Sznajd-Weron reports financial support was provided by National Science Centre Poland.

\section*{Acknowledgment}This research was funded by the National Science Center (NCN, Poland) through grant no. 2019\slash{}35\slash{}B\slash{}HS6\slash{}02530.

\section*{Copyright information} © 2022. This manuscript version is made available under the CC-BY-NC-ND 4.0 license https://creativecommons.org/licenses/by-nc-nd/4.0/

\FloatBarrier
 \bibliographystyle{elsarticle-num}

\begin{thebibliography}{10}
\expandafter\ifx\csname url\endcsname\relax
  \def\url#1{\texttt{#1}}\fi
\expandafter\ifx\csname urlprefix\endcsname\relax\def\urlprefix{URL }\fi
\expandafter\ifx\csname href\endcsname\relax
  \def\href#1#2{#2} \def\path#1{#1}\fi

\bibitem{Nai:Dom:Mac:13}
P.~R. Nail, S.~I.~D. Domenico, G.~MacDonald, Proposal of a double diamond model
  of social response, Review of General Psychology 17~(1) (2013) 1--19.
\newblock \href {https://doi.org/10.1037/a0030997}
  {\path{doi:10.1037/a0030997}}.

\bibitem{Mac:etal:21}
M.~W. Macy, M.~Ma, D.~R. Tabin, J.~Gao, B.~K. Szymanski, Polarization and
  tipping points, Proceedings of the National Academy of Sciences 118~(50)
  (2021) e2102144118.
\newblock \href {https://doi.org/10.1073/pnas.2102144118}
  {\path{doi:10.1073/pnas.2102144118}}.

\bibitem{Mas:Fla:13}
M.~Mäs, A.~Flache, Differentiation without distancing. explaining
  bi-polarization of opinions without negative influence, PLOS ONE 8~(11)
  (2013) 1--17.
\newblock \href {https://doi.org/10.1371/journal.pone.0074516}
  {\path{doi:10.1371/journal.pone.0074516}}.

\bibitem{Lan:Fou:Hol:18}
K.~Lanaj, T.~A. Foulk, J.~R. Hollenbeck, The benefits of not seeing eye to eye
  with leadership: Divergence in risk preferences impacts multiteam system
  behavior and performance, Academy of Management Journal 61~(4) (2018)
  1554--1582.
\newblock \href {https://doi.org/10.5465/amj.2015.0946}
  {\path{doi:10.5465/amj.2015.0946}}.

\bibitem{Min:Gin:22}
J.~A. Minson, F.~Gino, Managing a polarized workforce: How to foster debate and
  promote trust, Harvard Business Review 100~(2) (2022) 62--71.

\bibitem{Def:etal:02}
G.~Deffuant, F.~Amblard, G.~Weisbuch, T.~Faure, {How can extremism prevail? A
  study based on the relative agreement interaction model}, Journal of
  Artificial Societies and Social Simulation 5~(4) (2002).

\bibitem{Heg:Kra:02}
R.~Hegselmann, U.~Krause, {Opinion dynamics and bounded confidence: Models,
  analysis and simulation}, Journal of Artificial Societies and Social
  Simulation 5~(3) (2002).

\bibitem{Par:And:Mel:16}
A.~Parravano, A.~Andina-Díaz, M.~A. Meléndez-Jiménez, Bounded confidence
  under preferential flip: A coupled dynamics of structural balance and
  opinions, PLOS ONE 11~(10) (2016) 1--23.
\newblock \href {https://doi.org/10.1371/journal.pone.0164323}
  {\path{doi:10.1371/journal.pone.0164323}}.

\bibitem{Maa:Dal:Wal:20}
H.~L.~J. van~der Maas, J.~Dalege, L.~Waldorp, {The polarization within and
  across individuals: the hierarchical Ising opinion model}, Journal of Complex
  Networks 8~(2) (2020).
\newblock \href {https://doi.org/10.1093/comnet/cnaa010}
  {\path{doi:10.1093/comnet/cnaa010}}.

\bibitem{McP:Smi:Coo:01}
M.~McPherson, L.~Smith-Lovin, J.~M. Cook, Birds of a feather: Homophily in
  social networks, Annual Review of Sociology 27~(1) (2001) 415--444.
\newblock \href {https://doi.org/10.1146/annurev.soc.27.1.415}
  {\path{doi:10.1146/annurev.soc.27.1.415}}.

\bibitem{Lio:etal:17}
V.~Liordos, V.~J. Kontsiotis, M.~Georgari, K.~Baltzi, I.~Baltzi, Public
  acceptance of management methods under different human–wildlife conflict
  scenarios, Science of The Total Environment 579 (2017) 685--693.
\newblock \href {https://doi.org/10.1016/j.scitotenv.2016.11.040}
  {\path{doi:10.1016/j.scitotenv.2016.11.040}}.

\bibitem{Sta:Tes:Sch:08}
H.-U. Stark, C.~J. Tessone, F.~Schweitzer, Slower is faster: Fostering
  consensus formation by heterogeneous inertia, Advances in Complex Systems
  11~(04) (2008) 551--563.
\newblock \href {https://doi.org/10.1142/S0219525908001805}
  {\path{doi:10.1142/S0219525908001805}}.

\bibitem{Bis:Cha:Sen:12}
S.~Biswas, A.~Chatterjee, P.~Sen, Disorder induced phase transition in kinetic
  models of opinion dynamics, Physica A: Statistical Mechanics and its
  Applications 391~(11) (2012) 3257--3265.
\newblock \href {https://doi.org/10.1016/j.physa.2012.01.046}
  {\path{doi:10.1016/j.physa.2012.01.046}}.

\bibitem{Juu:Por:19}
J.~S. Juul, M.~A. Porter, Hipsters on networks: How a minority group of
  individuals can lead to an antiestablishment majority, Phys. Rev. E 99 (2019)
  022313.
\newblock \href {https://doi.org/10.1103/PhysRevE.99.022313}
  {\path{doi:10.1103/PhysRevE.99.022313}}.

\bibitem{Gra:Li:20}
M.~Grabisch, F.~Li, {Anti-conformism in the Threshold Model of Collective
  Behavior}, Dynamic Games and Applications 10 (2020) 444–477.
\newblock \href {https://doi.org/10.1007/s13235-019-00332-0}
  {\path{doi:10.1007/s13235-019-00332-0}}.

\bibitem{Axe:97}
R.~Axelrod, The dissemination of culture: A model with local convergence and
  global polarization, Journal of Conflict Resolution 41~(2) (1997) 203--226.
\newblock \href {https://doi.org/10.1177/0022002797041002001}
  {\path{doi:10.1177/0022002797041002001}}.

\bibitem{Mar:Men:14}
A.~Marvakis, M.~Mentinis, {Social Psychology}, 10th Edition, Springer New York,
  2014.
\newblock \href {https://doi.org/10.1007/978-1-4614-5583-7\_291}
  {\path{doi:10.1007/978-1-4614-5583-7\_291}}.

\bibitem{Bon:05}
R.~Bond, Group size and conformity, Group Processes \& Intergroup Relations
  8~(4) (2005) 331--354.
\newblock \href {https://doi.org/10.1177/1368430205056464}
  {\path{doi:10.1177/1368430205056464}}.

\bibitem{Nai:Mac:Lev:00}
P.~R. Nail, G.~MacDonald, D.~A. Levy, {Proposal of a four-dimensional model of
  social response}, Psychological Bulletin 126~(3) (2000) 454--470.
\newblock \href {https://doi.org/10.1037/0033-2909.126.3.454}
  {\path{doi:10.1037/0033-2909.126.3.454}}.

\bibitem{Cas:Mun:Pas:09}
C.~Castellano, M.~A. Mu\~noz, R.~Pastor-Satorras, Nonlinear $q$-voter model,
  Phys. Rev. E 80 (2009) 041129.
\newblock \href {https://doi.org/10.1103/PhysRevE.80.041129}
  {\path{doi:10.1103/PhysRevE.80.041129}}.

\bibitem{Jed:Szn:19}
A.~Jędrzejewski, K.~Sznajd-Weron, Statistical physics of opinion formation: Is
  it a spoof?, Comptes Rendus Physique 20~(4) (2019) 244--261.
\newblock \href {https://doi.org/10.1016/j.crhy.2019.05.002}
  {\path{doi:10.1016/j.crhy.2019.05.002}}.

\bibitem{Now:Sto:Szn:21}
B.~Nowak, B.~Stoń, K.~Sznajd-Weron, Discontinuous phase transitions in the
  multi-state noisy q-voter model: quenched vs. annealed disorder, Scientific
  Reports 11 (2021) 6098.
\newblock \href {https://doi.org/10.1038/s41598-021-85361-9}
  {\path{doi:10.1038/s41598-021-85361-9}}.

\bibitem{Now:Szn:22}
B.~Nowak, K.~Sznajd-Weron, Switching from a continuous to a discontinuous phase
  transition under quenched disorder, Phys. Rev. E 106 (2022) 014125.
\newblock \href {https://doi.org/10.1103/PhysRevE.106.014125}
  {\path{doi:10.1103/PhysRevE.106.014125}}.

\bibitem{Don:Lip:Szn:22}
M.~Doniec, A.~Lipiecki, K.~Sznajd-Weron, Consensus, polarization and hysteresis
  in the three-state noisy q-voter model with bounded confidence, Entropy
  24~(7) (2022) 983.
\newblock \href {https://doi.org/10.3390/e24070983}
  {\path{doi:10.3390/e24070983}}.

\bibitem{Dat:18}
G.~Datseris, Dynamicalsystems.jl: A {J}ulia software library for chaos and
  nonlinear dynamics, Journal of Open Source Software 3~(23) (2018) 598.
\newblock \href {https://doi.org/10.21105/joss.00598}
  {\path{doi:10.21105/joss.00598}}.

\bibitem{Fla:etal:17}
A.~Flache, M.~M\"{a}s, T.~Feliciani, E.~Chattoe-Brown, G.~Deffuant, S.~Huet,
  J.~Lorenz, Models of social influence: Towards the next frontiers, Journal of
  Artificial Societies and Social Simulation 20~(4) (2017) 2.
\newblock \href {https://doi.org/10.18564/jasss.3521}
  {\path{doi:10.18564/jasss.3521}}.

\bibitem{Kur:Mas:Lor:16}
T.~Kurahashi-Nakamura, M.~M\"{a}s, J.~Lorenz, Robust clustering in generalized
  bounded confidence models, Journal of Artificial Societies and Social
  Simulation 19~(4) (2016) 7.
\newblock \href {https://doi.org/10.18564/jasss.3220}
  {\path{doi:10.18564/jasss.3220}}.

\bibitem{Ban:Olb:19}
S.~Banisch, E.~Olbrich, Opinion polarization by learning from social feedback,
  The Journal of Mathematical Sociology 43~(2) (2019) 76--103.
\newblock \href {https://doi.org/10.1080/0022250X.2018.1517761}
  {\path{doi:10.1080/0022250X.2018.1517761}}.

\bibitem{Vaz:Kra:Red:03}
F.~Vazquez, P.~L. Krapivsky, S.~Redner, Constrained opinion dynamics: freezing
  and slow evolution, Journal of Physics A: Mathematical and General 36~(3)
  (2003) L61--L68.
\newblock \href {https://doi.org/10.1088/0305-4470/36/3/103}
  {\path{doi:10.1088/0305-4470/36/3/103}}.

\bibitem{Vaz:Red:04}
F.~Vazquez, S.~Redner, Ultimate fate of constrained voters, Journal of Physics
  A: Mathematical and General 37~(35) (2004) 8479--8494.
\newblock \href {https://doi.org/10.1088/0305-4470/37/35/006}
  {\path{doi:10.1088/0305-4470/37/35/006}}.

\bibitem{Mob:11}
M.~Mobilia, Fixation and polarization in a three-species opinion dynamics
  model, {EPL} (Europhysics Letters) 95~(5) (2011) 50002.
\newblock \href {https://doi.org/10.1209/0295-5075/95/50002}
  {\path{doi:10.1209/0295-5075/95/50002}}.

\bibitem{Cro:14}
N.~Crokidakis, A three-state kinetic agent-based model to analyze tax evasion
  dynamics, Physica A: Statistical Mechanics and its Applications 414 (2014)
  321--328.
\newblock \href {https://doi.org/10.1016/j.physa.2014.07.056}
  {\path{doi:10.1016/j.physa.2014.07.056}}.

\bibitem{Bal:Pin:Sem:15}
P.~Balenzuela, J.~P. Pinasco, V.~Semeshenko, The undecided have the key:
  Interaction-driven opinion dynamics in a three state model, PLOS ONE 10~(10)
  (2015) 1--21.
\newblock \href {https://doi.org/10.1371/journal.pone.0139572}
  {\path{doi:10.1371/journal.pone.0139572}}.

\bibitem{Rad:Don:21}
W.~Radosz, M.~Doniec, Three-state opinion q-voter model with bounded
  confidence, Lecture Notes in Computer Science 12744 (2021) 295--301.
\newblock \href {https://doi.org/10.1007/978-3-030-77967-2\_24}
  {\path{doi:10.1007/978-3-030-77967-2\_24}}.

\end{thebibliography}

\end{document}